     
\font\twelverm=cmr10 scaled 1200    \font\twelvei=cmmi10 scaled 1200
\font\twelvesy=cmsy10 scaled 1200   \font\twelveex=cmex10 scaled 1200
\font\twelvebf=cmbx10 scaled 1200   \font\twelvesl=cmsl10 scaled 1200
\font\twelvett=cmtt10 scaled 1200   \font\twelveit=cmti10 scaled 1200
\font\twelvesc=cmcsc10 scaled 1200  
\skewchar\twelvei='177   \skewchar\twelvesy='60
     
     
\def\twelvepoint{\normalbaselineskip=12.4pt plus 0.1pt minus 0.1pt
  \abovedisplayskip 12.4pt plus 3pt minus 9pt
  \belowdisplayskip 12.4pt plus 3pt minus 9pt
  \abovedisplayshortskip 0pt plus 3pt
  \belowdisplayshortskip 7.2pt plus 3pt minus 4pt
  \smallskipamount=3.6pt plus1.2pt minus1.2pt
  \medskipamount=7.2pt plus2.4pt minus2.4pt
  \bigskipamount=14.4pt plus4.8pt minus4.8pt
  \def\rm{\fam0\twelverm}          \def\it{\fam\itfam\twelveit}%
  \def\sl{\fam\slfam\twelvesl}     \def\bf{\fam\bffam\twelvebf}%
  \def\mit{\fam 1}                 \def\cal{\fam 2}%
  \def\sc{\twelvesc}               \def\tt{\twelvett}
  \def\sf{\twelvesf}
  \textfont0=\twelverm   \scriptfont0=\tenrm   \scriptscriptfont0=\sevenrm
  \textfont1=\twelvei    \scriptfont1=\teni    \scriptscriptfont1=\seveni
  \textfont2=\twelvesy   \scriptfont2=\tensy   \scriptscriptfont2=\sevensy
  \textfont3=\twelveex   \scriptfont3=\twelveex  \scriptscriptfont3=\twelveex
  \textfont\itfam=\twelveit
  \textfont\slfam=\twelvesl
  \textfont\bffam=\twelvebf \scriptfont\bffam=\tenbf
  \scriptscriptfont\bffam=\sevenbf
  \normalbaselines\rm}
     

     
\def\beginlinemode{\endmode
  \begingroup\parskip=0pt \obeylines\def\\{\par}\def\endmode{\par\endgroup}}
\def\beginparmode{\endmode
  \begingroup \def\endmode{\par\endgroup}}
\let\endmode=\par
{\obeylines\gdef\
{}}
\def\singlespace{\baselineskip=\normalbaselineskip}

\def\oneandahalfspace{\baselineskip=\normalbaselineskip
  \multiply\baselineskip by 3 \divide\baselineskip by 2}
\def\doublespace{\baselineskip=\normalbaselineskip \multiply\baselineskip by 2}

\newcount\firstpageno
\firstpageno=2
\footline={\ifnum\pageno<\firstpageno{\hfil}\else{\hfil\twelverm\folio\hfil}\fi}
\def\toppageno{\global\footline={\hfil}\global\headline
  ={\ifnum\pageno<\firstpageno{\hfil}\else{\hfil\twelverm\folio\hfil}\fi}}
\let\rawfootnote=\footnote              
\def\footnote#1#2{{\rm\singlespace\parindent=0pt\parskip=0pt
  \rawfootnote{#1}{#2\hfill\vrule height 0pt depth 6pt width 0pt}}}
\def\raggedcenter{\leftskip=4em plus 12em \rightskip=\leftskip
  \parindent=0pt \parfillskip=0pt \spaceskip=.3333em \xspaceskip=.5em
  \pretolerance=9999 \tolerance=9999
  \hyphenpenalty=9999 \exhyphenpenalty=9999 }
\def\dateline{\rightline{\ifcase\month\or
  January\or February\or March\or April\or May\or June\or
  July\or August\or September\or October\or November\or December\fi
  \space\number\year}}
\def\received{\vskip 3pt plus 0.2fill
 \centerline{\sl (Received\space\ifcase\month\or
  January\or February\or March\or April\or May\or June\or
  July\or August\or September\or October\or November\or December\fi
  \qquad, \number\year)}}
     
     
\hsize=6.5truein
\vsize=8.5truein
\parskip=\medskipamount
\def\\{\cr}
\twelvepoint            
\doublespace            
\overfullrule=0pt       

\def\title                      
  {\null\vskip 3pt plus 0.2fill
   \beginlinemode \doublespace \raggedcenter \bf}
     
\def\author                     
  {\vskip 3pt plus 0.2fill \beginlinemode
   \singlespace \raggedcenter\sc}
     
\def\affil                      
  {\vskip 3pt plus 0.1fill \beginlinemode
   \oneandahalfspace \raggedcenter \sl}
     
\def\abstract                   
  {\vskip 3pt plus 0.3fill \beginparmode
   \singlespace ABSTRACT: }
     
\def\endtopmatter               
  {\endpage                     
   \body}
     
\def\body                       
  {\beginparmode}               
     
\def\head#1{                    
  \goodbreak\vskip 0.5truein    
  {\immediate\write16{#1}
   \raggedcenter \uppercase{#1}\par}
   \nobreak\vskip 0.25truein\nobreak}

\def\beginitems{
\par\medskip\bgroup\def\i##1 {\item{##1}}\def\ii##1 {\itemitem{##1}}
\leftskip=36pt\parskip=0pt}
\def\enditems{\par\egroup}
     
\def\beneathrel#1\under#2{\mathrel{\mathop{#2}\limits_{#1}}}
     
\def\refto#1{$^{#1}$}           
     
\def\references                 
  {\head{References}            
   \beginparmode
   \frenchspacing \parindent=0pt \leftskip=1truecm
   \parskip=8pt plus 3pt \everypar{\hangindent=\parindent}}

\gdef\refis#1{\item{#1.\ }}                     
     
\gdef\journal#1, #2, #3, 1#4#5#6{               
    {\sl #1~}{\bf #2}, #3 (1#4#5#6)}            

\gdef\refa#1, #2, #3, #4, 1#5#6#7.{\noindent#1, #2 {\bf #3}, #4 (1#5#6#7).\rm} 

\gdef\refb#1, #2, #3, #4, 1#5#6#7.{\noindent#1 (1#5#6#7), #2 {\bf #3}, #4.\rm} 

\def\pr{\journal Phys.Rev., }

\def\endreferences{\body}

\def\endpage                    
  {\vfill\eject}
     
\def\endpaper                   
  {\endmode\vfill\supereject}

\def\ref#1{Ref.~#1}                     
\def\Ref#1{Ref.~#1}                     
\def\[#1]{[\cite{#1}]}
\def\cite#1{{#1}}
\def\(#1){(\call{#1})}
\def\call#1{{#1}}
\def\taghead#1{}
\def\frac#1#2{{#1 \over #2}}
\def\half{{\frac 12}}

\def\12{{1\over2}}

\catcode`@=11
\newcount\r@fcount \r@fcount=0
\newcount\r@fcurr
\immediate\newwrite\reffile
\newif\ifr@ffile\r@ffilefalse
\def\w@rnwrite#1{\ifr@ffile\immediate\write\reffile{#1}\fi\message{#1}}

\def\writer@f#1>>{}
\def\referencefile{
  \r@ffiletrue\immediate\openout\reffile=\jobname.ref%
  \def\writer@f##1>>{\ifr@ffile\immediate\write\reffile%
    {\noexpand\refis{##1} = \csname r@fnum##1\endcsname = %
     \expandafter\expandafter\expandafter\strip@t\expandafter%
     \meaning\csname r@ftext\csname r@fnum##1\endcsname\endcsname}\fi}%
  \def\strip@t##1>>{}}

\def\citeall#1{\xdef#1##1{#1{\noexpand\cite{##1}}}}
\def\cite#1{\each@rg\citer@nge{#1}}	

\def\each@rg#1#2{{\let\thecsname=#1\expandafter\first@rg#2,\end,}}
\def\first@rg#1,{\thecsname{#1}\apply@rg}	
\def\apply@rg#1,{\ifx\end#1\let\next=\relax
\else,\thecsname{#1}\let\next=\apply@rg\fi\next}

\def\citer@nge#1{\citedor@nge#1-\end-}	
\def\citer@ngeat#1\end-{#1}
\def\citedor@nge#1-#2-{\ifx\end#2\r@featspace#1 
  \else\citel@@p{#1}{#2}\citer@ngeat\fi}	
\def\citel@@p#1#2{\ifnum#1>#2{\errmessage{Reference range #1-#2\space is bad.}%
    \errhelp{If you cite a series of references by the notation M-N, then M and
    N must be integers, and N must be greater than or equal to M.}}\else%
 {\count0=#1\count1=#2\advance\count1 by1\relax\expandafter\r@fcite\the\count0,
  \loop\advance\count0 by1\relax
    \ifnum\count0<\count1,\expandafter\r@fcite\the\count0,%
  \repeat}\fi}

\def\r@featspace#1#2 {\r@fcite#1#2,}	
\def\r@fcite#1,{\ifuncit@d{#1}
    \newr@f{#1}%
    \expandafter\gdef\csname r@ftext\number\r@fcount\endcsname%
                     {\message{Reference #1 to be supplied.}%
                      \writer@f#1>>#1 to be supplied.\par}%
 \fi%
 \csname r@fnum#1\endcsname}
\def\ifuncit@d#1{\expandafter\ifx\csname r@fnum#1\endcsname\relax}%
\def\newr@f#1{\global\advance\r@fcount by1%
    \expandafter\xdef\csname r@fnum#1\endcsname{\number\r@fcount}}

\let\r@fis=\refis			
\def\refis#1#2#3\par{\ifuncit@d{#1}
   \newr@f{#1}%
   \w@rnwrite{Reference #1=\number\r@fcount\space is not cited up to now.}\fi%
  \expandafter\gdef\csname r@ftext\csname r@fnum#1\endcsname\endcsname%
  {\writer@f#1>>#2#3\par}}

\def\ignoreuncited{
   \def\refis##1##2##3\par{\ifuncit@d{##1}%
    \else\expandafter\gdef\csname r@ftext\csname r@fnum##1\endcsname\endcsname%
     {\writer@f##1>>##2##3\par}\fi}}

\def\r@ferr{\endreferences\errmessage{I was expecting to see
\noexpand\endreferences before now;  I have inserted it here.}}
\let\r@ferences=\references
\def\references{\r@ferences\def\endmode{\r@ferr\par\endgroup}}

\let\endr@ferences=\endreferences
\def\endreferences{\r@fcurr=0
  {\loop\ifnum\r@fcurr<\r@fcount
    \advance\r@fcurr by 1\relax\expandafter\r@fis\expandafter{\number\r@fcurr}%
    \csname r@ftext\number\r@fcurr\endcsname%
  \repeat}\gdef\r@ferr{}\endr@ferences}


\let\r@fend=\endpaper\gdef\endpaper{\ifr@ffile
\immediate\write16{Cross References written on []\jobname.REF.}\fi\r@fend}

\catcode`@=12

\citeall\refto		
\citeall\ref		%
\citeall\Ref		%

\def\refto#1{$^{#1}$}         

\def\la{\langle}
\def\ra{\rangle}
\def\ria{\rightarrow}

\def\x{{\bf x}}

\def\k{{\bf k}}
\def\q{{\bf q}}

\def\a{\alpha}

\def\s{{\sigma}}

\def\ih{{i \over \hbar}}
\def\au{{\underline{\alpha}}}

\def\s{{\sigma}}

\def\p{{\bf p }}

\centerline{\bf The Emergence of Hydrodynamic Equations from Quantum Theory:} 
\centerline{\bf A Decoherent Histories Analysis}

\author J.J.Halliwell 
\affil 
Theory Group 
Blackett Laboratory 
Imperial College 
London 
SW7 2BZ 
UK 
\vskip 0.5in 
\centerline {\rm Preprint Imperial/TP/99-0/14. December, 1999.} 
\vskip 0.5in  
\centerline {\rm To appear in Proceedings of the 4th Peyresq Conference, 1999.} 
\vskip 1.0in 

\abstract{The most general description of the classical world is in
terms of local densities (such as number, momentum, energy), and
these typically evolve according to evolution equations of
hydrodynamic form. To explain the emergent classicality of these
variables from an underlying quantum theory, it is therefore
necessary to show, firstly, that these variables exhibit negligible
interference, and secondly, that the probabilities for histories of
them are peaked around hydrodynamic evolution. The
implementation of this programme in the context of the decoherent
histories approach to quantum theory is described. It is argued
that, for a system of weakly interacting particles, the eigenstates
of local densities (averaged over a sufficiently large volume)
remain approximate eigenstates under time evolution. This is a
consequence of their close connection with the corresponding
exactly conserved (and so exactly decoherent) quantities. The
subsequent derivation of hydrodynamic equations from decoherent
histories is discussed.}

\endtopmatter 
\endpage

If the universe is described at the most fundamental level by 
quantum theory, why is it so very nearly classical? There  are very
many aspects to the issue of emergent classicality (see, for
example, Ref.[\cite{Har6}] for an overview), but crucial to most of
them is the demonstration that certain types of quantum states of
the system in question exhibit negligible interference.  Initial
superpositions of such states may therefore be effectively replaced
by statistical mixtures. This, loosely speaking, is decoherence, and
has principally been demonstrated for the situation in which there
is a distinguished system, such as a particle, coupled to its
surrounding environment [\cite{JoZ,Zur}].

Most generally, decoherence typically comes about when the variables
describing the entire system of interest naturally separate into
``slow'' and ``fast'', whether or not this separation corresponds
to, respectively, system and environment\footnote{$^{\dag}$}{See
Ref.[\cite{BrHa}] for a discussion of the  conditions under which
the total Hilbert space may be  written as a tensor product of
system and environment Hilbert spaces}. If the system consists of a
large collection of interacting identical particles, as in a fluid
for example, the natural set of slow variables are the local
densities: energy, momentum, number, charge {\it etc.}.  These
variables, in fact, are also the variables which provide the most
complete description of the {\it classical} state of a fluid at a
macroscopic level. 

The most general demonstration of emergent
classicality therefore  consists of showing that, for a large
collection of interacting particles described microscopially by
quantum theory, the local densities become effectively classical.
Although one might argue that the system--environment mechanism
might play a role, since the collection of particles are coupled to
each other, decoherence comes about in these situations for a
different reason: it is because the local densities are almost
conserved if averaged over a sufficiently large volume
[\cite{GH1}]. Hence, the approximate non-interference of
local densities is due to the fact that they are close to a set of
exactly conserved quantities, and exactly conserved quantities obey
superselection rules.

Intuitively appealing though this argument is, it is clearly a {\it
quantitative} issue. The object of this letter is to show that,
under certain reasonable conditions,  local densities averaged over
a sufficiently large volume are indeed  approximately decoherent as
a result of their close connection to exact conservation.

We will approach the question using the decoherent histories
approach to quantum theory [\cite{GH1,Gri,Omn,Hal1}], which
has proved particularly useful for discussing emergent classicality
in a variety of contexts.
The central object of interest is the decoherence functional,
$$
\eqalignno{
D (\au, \au' ) = & {\rm Tr} \left( P_{\a_n} 
e^{-\ih H (t_n - t_{n-1} )} \cdots 
P_{\a_2} e^{-\ih H (t_2 - t_1)} P_{\a_1} |\Psi \ra \right.
\cr & \left. \times
\la \Psi | P_{\a_1'} e^{\ih H (t_2 - t_1)} 
P_{\a_2'} \cdots P_{\a_{n-1}' } 
e^{\ih H (t_n - t_{n-1} )}
\right)
& (1) \cr}
$$
The histories are characterized by the initial state $ | \Psi \ra $
and by the strings of projection operators $P_{\a}$ at times $t_1$
to $t_n$ (and $\au$ denotes the string of alternatives $\a_1 \cdots
\a_n$). Intuitively, the decoherence functional is a measure of the
interference between pairs of histories $\au$, $\au'$. When it is
zero for $\au \ne \au' $, we say that the histories are decoherent and
probabilities  $ p (\au ) = D (\au, \au ) $ obeying the usual 
probability sum rules may be assigned to them. 
One can then ask whether these
probabilities are strongly peaked about trajectories obeying
classical equations of motion. For the local densities, these
equations will be hydrodynamic equations, and these and closely
related aspects of emergent classicality have been pursued at
greater length elsewhere Refs.[\cite{BrHa,Ana,Hal2}]. 

We consider the class of systems which are described at the
microscopic level by a Hamiltonian of the form
$$
H = \sum_j \left( { {\bf p}_j^2 \over 2 m } + \sum_{\ell >j } \phi
( | \q_j - \q_\ell | ) \right)
\eqno(2)
$$
For definiteness, we will concentrate on the case of a weakly
interacting dilute gas, making brief reference to a one-dimensional
chain of oscillators, but it will be clear that the physical
ideas are reasonably general.
The local densities of interest are the number density
$n(\x)$, the momentum density ${\bf g}(\x)$ and the energy
density $ h (\x )$, defined by,
$$
\eqalignno{
n(\x) &= \sum_j \ \delta(\x -\q_j)
&(3) \cr
{\bf g}(\x) &= \sum_j \ {\bf p}_j \ \delta (\x- \q_j)
&(4) \cr
h(\x) &= \sum_j \ \left( { {\bf p}^2_j \over 2 m } + \sum_{\ell > j}
\phi  (| \q_j - \q_{\ell} |) \right)\ \delta(\x- \q_j)
&(5) \cr }
$$
(suitably ordered, in the quantum case).
We are interested in local densities smeared over a volume $V$.
The effect of this is to replace the delta functions with a window
function, denoted $\delta_V$, which is zero outside $V$ and $1$
inside. It is also 
useful to work with the Fourier transforms of the local densities,
denoted $n (\k) $, $g (\k) $, $h(\k )$. So, for example,
the local number density at wavelength $\k$ is
$$
n(\k) = \sum_j \ e^{i \k \cdot \q_j}
\eqno(6)
$$
Exact conservation is obtained in the limit $ k = | \k | \ria  0 $, 
or $V \ria \infty $ in (3)--(5).

We would like to compute the decoherence functional for histories
consisting of projections onto the operators (3)--(5). (The
construction of the projectors is described in more detail in
Ref.[\cite{Hal2}]). In the case of exact conservation, $k=0$,  we
have exact decoherence simply because the projectors in
Eq.(1) all commute with $H$ and with each other [\cite{HLM}]. Our
main task is therefore to show that as $k$ increases from zero there
is still a non-trivial regime in which decoherence is approximately
maintained.

A significant result in this direction has been established already
by Calzetta and Hu for the case of local temperature $T(x)$ obeying
the diffusion equation [\cite{CaH}]. They took their initial state
to be close to the equilibrium state, and worked backwards from the
diffusion equation plus fluctuations to deduce the influence
functional it must have arisen from, from which the degree of
decoherence could be deduced. Here, by contrast, initial macroscopic
superposition states are considered. A more detailed comparison of
these two approaches is certainly of interest.

We begin by rewriting the exact conservation case in a simple way
that makes its generalization to locally conserved quantities more
apparent. Suppose the histories are projections onto some conserved
quantity, $Q$. Let the initial state be a superposition of
eigenstates of $Q$,
$$
| \Psi \ra = { 1 \over \sqrt{2} } \left( |a \ra + |b \ra \right)
\eqno(7)
$$
where $ \la a | b \ra = 0 $ and
$$
\hat Q | a \ra = a | a \ra, \quad \hat Q | b \ra = b | b \ra
\eqno(8)
$$
Since the $P_{\a}$'s are projections onto $Q$, $P_{\a}$
either annihilates or preserves
$ | a \ra $ and $ | b \ra $.
Take the case of a history with just two
moments of time (the generalization to more times is trivial).
The only non-zero off-diagonal terms of the decoherence functional 
are of the form
$$
\eqalignno{
D(\au, \au') 
&= \half {\rm Tr} \left( P_{\a_2} e^{ -\ih H t} | a \ra \la b | e^{
\ih H t } \right)
\cr
& = \half {\rm Tr} \left( P_{\a_2} | a_t \ra \la b_t | \right)
&(9)  \cr }
$$
But $Q$ is conserved, hence $[P_{\a_2},H] = 0 $ and
$$
\eqalignno{
P_{\a_2} | a_t \ra & = P_{\a_2 } e^{ - \ih H t } | a \ra 
\cr
& = e^{ - \ih H t } P_{\a_2} | a \ra = | a_t \ra
&(10) \cr }
$$ 
(or equals zero if $\a_2$ does not correspond to $a$).
It follows that
$$
\eqalignno{
D(\au, \au') & =
\half {\rm Tr} \left( P_{\a_2} | a_t \ra \la b_t  | \right) 
\cr
& = \la b_t | a_t \ra = \la b | a \ra = 0 
&(11) \cr} 
$$
and therefore we have decoherence.

Now suppose that the operator $Q$ is one of the local densities
(3)--(5), so is no longer exactly conserved. The steps up to Eq.(9)
still hold. But to go further, we need to know how the eigenstates
of the local densities behave under time evolution. A reasonable
supposition, which will be justified, is the following.
Let us suppose that under time evolution, the
eigenstates of $Q$ remain approximate eigenstates. That is, we
initially have (7), but under evolution to time $t$,
$$
\hat Q  | a_t \ra \approx \la Q \ra | a_t \ra 
\eqno(12)
$$
or, more precisely, 
$$
{ \left( \Delta Q \right)^2 \over \la Q  \ra^2 } << 1
\eqno(13)
$$
{\it i.e.}, the state remains strongly peaked in the variable $Q$
under time evolution. The states are then approximate eigenstates
of the projectors, so that in place of Eq.(10), we have 
the approximate result, $ P_{\a_2} | a_t \ra \approx | a_t \ra $
(or equals zero) as long as
the width of the projection is much
greater than the  uncertainty $ (\Delta Q)^2 $. Hence Eq.(11)
follows approximately, and we get approximate decoherence to the
extent that the approximation (13) holds.

The key point is therefore the following: approximate decoherence is
assured for histories of operators $Q$ whose eigenstates
have the property that they remain strongly
peaked in $Q$ under time evolution, as characterized by (13). To
demonstrate decoherence of the local densities, therefore, we need
only find their eigenstates, and show that they satisfy the
localization property (13) under time evolution. (Note, incidently,
that the above argument actually assures decoherence of {\it any}
variables $Q$ satisfying the localization property. The particular
significance of the local densities is that they are continuous
functions of the coarse graining scale $k$, so are guaranteed to
satisfy the requisite property if $k$ is sufficiently close to
zero.)

Since the three operators (3)--(5)
do not commute, exact simultaneous eigenstates cannot be found. However,
there are approximate
simultanous eigenstates. For weak interactions, 
they are products of $N$ identical terms,
$$ 
| \Psi \ra = | \psi \ra \otimes | \psi \ra \otimes 
\cdots \otimes | \psi \ra
\eqno(14) 
$$ 
and are approximate eigenstates of all three operators for large
$N$.  The proof of this statement involves considering, for the
local number density for example, the object $ ( \Delta n (\x) )^2 /
\la n (\x) \ra^2 $, and showing that it goes like $1/N$ for large
$N$ (see Ref.[\cite{Hal2}], for example).  It is essentially the
central limit theorem (see also Ref.[\cite{FGH}]). For the number
and momentum density it relies on the fact that they are sums of
identical one-particle operators. For the local energy density, it
additionally requires the smearing volume to be sufficiently large,
compared to some lengthscale indicated by the interactions. Some
tuning of the state $ | \psi \ra $ can be carried out to ensure that
(14) is an optimal approximate eigenstate of all the  local
densities but this will not be done here. (Also,  the passage to
exact eigenstates of $n(\k )$, $g( \k ) $, $h ( \k ) $  as $k \ria 0
$ can be seen explicitly if the one-particle states  $ | \psi \ra$
are taken to be one-particle momentum eigenstates).

The question is now what happens to the eigenstates (14) of the local
densities under time evolution by the Hamiltonian (2). Consider
first the trivial but enlightening case in which there no
interactions. In this case, the time evolved eigenstates $ | a_t
\ra $ remain of the product form (14), so they are {\it still}
approximate eigenstates of the local densities (but with a
time-evolved eigenvalue) for the same reasons as above. 
Hence there is approximate decoherence.

Decoherence in the non-interacting case comes about for two reasons.
First, it is due to the fact that a state of the form (14) will
remain strongly peaked about the average values of the local
densities, $n(\x)$, $g( \x ) $, $h (\x ) $ under time evolution, and
thus the state is essentially undisturbed by the projectors (as long
as their widths are sufficiently large). The strong peaking follows
from the assumption of large $N$ and from the fact that the local density
operators are sums of identical one particle operators. Secondly, it
is due to the almost trivial fact that the orthogonality of the two
elements of the initial state is preserved by unitary evolution.

This second fact is important because the first one is not always
sufficient to guarantee decoherence. Although the state remains
strongly peaked about the average values of the local densities,
these average values do not necessarily obey deterministic
equations. In the case of histories characterized by number density
only, for example, $ \la n (\x) \ra $ at time $t$ is {\it not}
uniquely determined by  $ \la n (\x ) \ra $ at the initial time (in
the state (14)). That is, in Eq.(9), $ | a_t \ra $ and   $ |b_t \ra
$ may in fact be peaked about {\it the same} value of number
density, even though the initial values are different.  The
decoherence is therefore not in fact due to an approximate 
determinism (such as that used in the phase space histories of
Omn\`es [\cite{Omn}]).  It is necessary only that the evolved states
are essentially undisturbed by the projectors, and therefore that
the two orthogonal components of the initial state are eventually
overlapped at the final time, as in Eq.(11), to give zero.

The next and most important task is to show that the above story is
in fact still true, with qualifications, in the presence of
interactions. The complete description of $N$ interacting particles
is generally extremely involved, but we can make some progress by
restricting attention to a sufficiently dilute gas of weakly
interacting components, and then making two assumptions which are
standard in kinetic theory and non-equilibrium statistical mechanics
[\cite{SM}]. It is notationally convenient in what follows to work
with a Wigner function, rather than quantum state. Hence associated
with the full $N$--particle wave function is an $N$--particle Wigner
function $ W_N (\p_1, \q_1, \cdots \p_N, \q_N ) $. For a dilute,
weakly interacting gas, it is reasonable to assume that
three--particle correlations are negligible. This is our first
assumption. It means that all the physics is contained in the one
and two--particle reduced Wigner functions, $W_1 (\p_1, \q_1)$ and $
W_2 (\p_1, \q_1, \p_2, \q_2 ) $. All higher order  reduced Wigner
functions will reduce to products of these.

We again take as our initial state the
product state (14) (which is still an approximate eigenstate in the
interacting case), and let it evolve, so correlations will develop.
The degree to which the particles become
correlated is contained in the two--particle distribution
$W_2$ of the evolved eigenstate. 
On general grounds, we expect that the inter-particle
correlations will only be important on some length scale $L$, and
beyond that length scale, they will be uncorrelated. That is,  we
will assume that
$$
W_2 (\p_1, \q_1, \p_2, \q_2 ) \approx W_1 (\p_1, \q_1 ) W_1 (\p_2,
\q_2 )
\eqno(15) 
$$
for $ | \q_2 - \q_1 | > L$, and otherwise $W_2$ will have a form
indicating non-trivial correlations. This is our second assumption.
It is physically reasonable, and it is in fact a key assumption in
the derivation of the Boltzmann equation [\cite{SM}].

Note that the assumption (15) would not necessarily be appropriate
for all possible initial quantum states. One could construct initial
quantum states which would possess or develop  non-trivial
long-range correlations, for which this assumption may never hold.
However, as seen in the non-interacting case, to demonstrate
decoherence we only need to consider the  time evolution of the
special class of initial states which are  eigenstates of the local
densities. These particular initial states, which are of the form
(14), approximately,  do not have long-range correlations. It is
therefore very plausible, at least for a dilute, weakly interacting
gas, that they will develop only limited correlations under time
evolution and the assumption (15) will hold.

Given the above assumptions, it is now reasonably straighforward to
argue that the state is still strongly peaked about the average values
of the local densities, as long as $V >> L^3 $. 
For example, for the number density, we have
$$
\la n(\x ) \ra = \sum_j \la \delta_V ( \q_j - \x ) \ra
= N \int_V d^3 \q \ p (\q )
\eqno(16) 
$$
where $ p ( \q ) $ is the one-particle 
probability distribution of $\q$ (obtained by integrating the
one-particle Wigner function over $\p $).
Similarly,
$$
\eqalignno{
\la n^2 (\x) \ra &= \sum_{j \ell} \ \la \delta_V (\q_j -\x ) \delta_V
(\q_{\ell} -\x ) \ra
\cr
&= N \la \delta_V \ra + (N^2 - N) \la \delta_V (\q_1 -x ) \delta_V
(\q_2 - x ) \ra
&(17) \cr}
$$
where we have used $\delta_V^2 = \delta_V $, and also an assumption
of identical particles to reduce the sum over $j,\ell$ to particles
labeled $1$ and $2$.
We now have
$$
\eqalignno{
( \Delta n (\x) )^2 
=& \la n^2 (\x)  \ra - \la n (\x) \ra^2 
\cr
= & N^2 \left( \la \delta_V (\q_1 - \x ) \delta_V (\q_2 - \x ) \ra
- \la \delta_V \ra^2 \right)
\cr
& + N \left( \la \delta_V \ra - \la \delta_V (\q_1 -\x ) \delta_V
(\q_2 - \x ) \ra \right)
&(18) \cr }
$$
If there is no correlation at all between the particles, the
coefficient
of $N^2$ would vanish, so $( \Delta n (\x) )^2 / \la n(\x )
\ra^2 $ would go like $ 1/ N $, which goes to zero as $N \ria
\infty$.
This is the standard central limit theorem result indicated earlier
for the non-interacting case. With interactions, the coefficient
of $N^2$ is no longer zero. We now need to show, therefore, that
this term
is still sufficiently small for  
$( \Delta n (\x) )^2 / \la n(\x ) \ra^2 $ 
to remain small as $N \ria \infty $. Introducing the
two-particle distribution $ p (\q_1, \q_2) $ (obtained by
integrating $\p_1, \p_2 $ out of $W_2$), it is readily shown that
the leftover terms as $N \ria \infty$ are
$$
{(\Delta n (\x) )^2 \over  \la n(\x )\ra^2 }
= { \int_V d^3 \q_1 \int_V d^3 \q_2 \left( p(\q_1, \q_2 )
- p(\q_1 ) p (\q_2 ) \right)
\over \left( \int_V d^3 \q \ p (\q ) \right)^2 }
\eqno(19)
$$
This is clearly zero if there are no correlations.
In the interacting case we use the assumption (15), which implies
that
$$
p(\q_1, \q_2 ) \approx p(\q_1 ) p (\q_2 )
\eqno(20)
$$
for $ | \q_1 - \q_2 | > L $, and otherwise non-trivial correlations
exist. Hence the integral in the numerator takes contributions
only from the region $ | \q_1 - \q_2 | < L $.

To see that (19) is small, note that in the numerator, the
integral is over a volume $V^2$ in the six-dimensional two particle
configuration space. If $ V << L^3 $, the factorization of  $ p
(\q_1, \q_2 ) $ for $ | \q_1 - \q_2 | > L $ makes no difference,
since $\q_1$ and $\q_2$ can never be far enough apart in the 
integrand (assuming $V$ is regular in shape). 
However, if $ V >> L^3 $, the $V^2$-sized integration
region is substantially reduced in size to $ V \times L^3 $. On dimensional
grounds the numerator is thereofre proportional to a number of order $ V L^3
$, and the denominator to $V^2 $ (perhaps with other factors common
to both). This means that
$$
{(\Delta n (\x) )^2 \over  \la n(\x )\ra^2 } \sim { L^3 \over V }
\eqno(21) 
$$
This order of magnitude estimate becomes exact if we assume
that the probabilities are constant in the region of non-trivial
correlation (another common assumption of kinetic theory [\cite{SM}]).
Hence the state will be strongly peaked about the average of
$ n(\x )$ if $ V >> L^3$.

In the one-dimensional oscillator chain model considered in 
Ref.[\cite{Hal3}], the uncertainty in $n(k)$ (the one-dimensional
version of Eq.(6)), can be computed explicitly in the special case
of a Gaussian state. It is, 
$$ 
( \Delta n ( k ) )^2 = \sum_{j=1}^N
\sum_{\ell=1}^N \la e^{ i k q_j } \ra \la e^{ - ik q_{\ell} } \ra
\left( e^{ k^2 \s (q_j, q_{\ell} )} -1 \right) 
\eqno(22) 
$$ 
where
$\s (q_j, q_{\ell} )  = \la q_j q_{\ell} \ra  - \la q_j \ra \la
q_{\ell} \ra $ measures the degree of correlation between different
particles in the chain. As $k$ increases from zero, the leading
order terms in (22) are of the form, $( \Delta n ( k ) )^2 = k^2
(\Delta X)^2 $, where $ X = \sum_j q_j $ (the centre of mass
coordinate), and since $ \la n(k) \ra \sim N $, we have
$$
{ ( \Delta n ( k ) )^2 \over | \la n (k) \ra |^2 } 
\sim { k^2 ( \Delta X)^2 \over  N^2 } 
\eqno(23)
$$ 
This will be very small as long as $k^{-1}$ is much larger than the
lengthscale of a single particle. $ ( \Delta n ( k ) )^2 $ 
starts to grow very rapidly with $k$, and (23) is no longer valid,
when $k^{-1}$
becomes less than the correlation length indicated by $\s (q_j,
q_{\ell} ) $.
Hence the state is strongly peaked about the mean as long as the 
coarse graining lengthscale $k^{-1}$ remains much greater than the
correlation length of the time-evolved local density eigenstates.
This correlation length is considered in Ref.[\cite{Hal3}] and found
to be generally very small compared to the system size. A simple
field theory model is also considered in Ref.[\cite{Hal3}],
confirming many of the expected features outlined here.

It is possible to see on physical grounds why one expects a result
of the form (21) to hold quite generally. In the non-interacting
case we used the  central limit theorem result that $ ( \Delta n )^2
/ \la n \ra^2 $ goes like $1/N$.  In the interacting case, the state
is no longer of the product form (14), but an analagous result still
holds. The point is that the correlations that develop extend only
over a (typically small) volume of size $L^3$, so the system breaks
up into a large number of essentially identical uncorrelated regions
of this size. Therefore each smearing volume $V$, if much greater
than $L^3$, contains of order $V/L^3$ identical  uncorrelated
regions each of which contribute equally to the local density
averaged over $V$. Loosely speaking, a central limit theorem-type
result again applies, not to the $N$ uncorrelated particles in the
same state, but to the  $ V/L^3 $ uncorrelated regions. So $1/N$ is
replaced by $L^3 /V $ in the central limit theorem, and hence the
above result.

Similar results hold for the local momentum and energy density. We
have therefore demonstrated the desired result: eigenstates of the
coarse-grained local densities  remain approximate eigenstates under
time evolution as long as the smearing volume is much greater than
the correlation volume of these states. Decoherence of these
variables then follows.

We now briefly consider the probabilities for histories.  They are
strongly peaked at each moment of time about the average values, $
\la n (\x,t ) \ra $, $ \la g ( \x ,t) \ra $,  $ \la h ( \x,t) \ra $,
averaged in a local density eigenstate.  The hydrodynamic equations
(or even a closed set of equations) do not necessarily follow,
however, since these require a local equilibrium initial state
[\cite{For,SM}], but we outline how this might come about.

The averages of the local densities depend only on the one-particle
Wigner function (with a small correction depending on the two
particle function in the case of local energy density), hence the
evolution equation of the local densities can be determined by
obtaining an evolution equation for the one-particle Wigner 
function $W_1$. With a local density eigenstate as initial state,
and with the two assumptions utilized above,  we expect this
evolution equation is the Boltzmann equation. We therefore evolve
$W_1$, subject to the initial condition that it be equal to the
one-particle Wigner function of a local density eigenstate. This
initial Wigner function is not of local equilibrium form, but it is
reasonable to expect that  it will rapidly approach local
equilibrium form under evolution according to the Boltzmann
equation, and thereafter retain that form. (The temperature,
chemical potential {\it etc.}, of the local equilibrium state will
be determined by the average values $ \la h ( \x,t) \ra $ etc.).
Hence, except for a short initial period during which the initial
state settles down to local equilibrium form, the probabilities for
histories will be peaked about hydrodynamic equations.

Some of these features can be seen in some detail in
Ref.[\cite{Hal2}], where the emergence of the diffusion equation 
was considered. The system studied was a collection of $N$ foreign
non-interacting particles in a background fluid. Decoherence was
therefore provided largely by the fluid in this case, rather than by
conservation, but the interest of the model is that it gives an
explicit picture of the emergence of a hydrodynamic equation. Each
foreign particle behaves like a quantum Brownian particle, whose
evolution equation is well-known. An initial state of the form (14)
for the $N$ foreign particles evolves into a mixed state of the form
of an $N$-fold product,
$$
\rho = \rho_1 \otimes \rho_1 \cdots \otimes \rho_1
\eqno(24)
$$
where each $1$-particle density operator $\rho_1$ describes quantum
Brownian motion. From the Wigner function of $\rho_1$ it is readily
shown that the $1$-particle position distribution obeys the
diffusion equation at long times, from which it readily follows
that the $N$-particle number density $n(x)$ also obeys the diffusion
equation. This model is therefore a kind of ``half-way house''
between the decoherence-through-environment and 
decoherence-through-conservation models, but it helps to complete
the general picture.

Summarizing, the final picture we have is therefore as follows.
An initial state consisting of a superposition of local
density eigenstates may be treated as a mixture of the same
states, since they are decoherent. Each state separately
will give probabilities peaked about hydrodynamic equations,
with particular values of initial values of phenomenological
parameters such as temperature {\it etc.}, and these will be different for
each element of the mixture. We therefore have a statistical mixture
of trajectories, each evolving according to hydrodynamic equations
but with different phenomenological parameters, {\it i.e.}, to very 
different macroscopic states. More details of this work may be found
elsewhere [\cite{Hal3,Hal4}].

\head{\bf Acknowledgements}

I am grateful to Todd Brun, Jim Hartle, Ray Rivers and Tom Kibble
for useful conversations. I would particularly like the thank the
organizers of Peyresq 4, Edgard Gunzig and Enric Verdaguer, for
inviting me to take part.  Thanks also to the local staff in
Peyresq for their hospitality.

\references
\def\pr{{\sl Phys.Rev.}}

\refis{Ana} C.Anastopoulos,
preprint gr-qc/9805074 (1998);
T.Brun and J.J.Halliwell, 
{\sl Phys.Rev.} {\bf 54}, 2899 (1996);
E. Calzetta and B. L. Hu, in {\it Directions in General
Relativity}, edited by B. L. Hu and T. A. Jacobson (Cambridge
University Press, Cambridge, 1993).

\refis{BrHa} T.Brun and J.B.Hartle, quant-ph/9905079.

\refis{CaH} E.A.Calzetta and B.L.Hu, {\sl Phys.Rev.} {\bf D59},
065018 (1999).

\refis{FGH} D.Finkelstein, {\sl Trans.N.Y.Acad.Sci.} {\bf 25}, 
621 (1963);
N.Graham, in {\it The Many Worlds Interpretation of
Quantum Mechanics}, B.S.DeWitt and N.Graham (eds.) (Princeton
University Press, Princeton, 1973);
J.B.Hartle, {\sl Am.J.Phys.} {\bf 36}, 704 (1968).
See also, E.Farhi, J.Goldstone and S.Gutmann, 
{\sl Ann.Phys.(NY)} {\bf 192}, 368 (1989).


\refis{For} D. Forster, {\it Hydrodynamic Fluctuations, Broken
Symmetry and Correlation Functions} (Benjamin, Reading, MA, 1975).

\refis{GH1} M.Gell-Mann and J.B.Hartle, in {\it Complexity, Entropy 
and the Physics of Information, SFI Studies in the Sciences of Complexity},
Vol. VIII, W. Zurek (ed.) (Addison Wesley, Reading, 1990);
{\sl Phys.Rev.} {\bf D47}, 3345 (1993).


\refis{Gri} R.B.Griffiths, {\sl J.Stat.Phys.} {\bf 36}, 219 (1984);
{\sl Phys.Rev.Lett.} {\bf 70}, 2201 (1993).

\refis{Hal1} J.J.Halliwell, in {\it Fundamental Problems in Quantum
Theory},  edited by D.Greenberger and A.Zeilinger, Annals of the New
York Academy of Sciences, Vol 775, 726 (1994).

\refis{Hal2} J.J.Halliwell, {\sl Phys.Rev.} {\bf D58}, 105015 (1998).

\refis{Hal3} J.J.Halliwell, in preparation.

\refis{Hal4} J.J.Halliwell, {\sl Phys.Rev.Lett.} {\bf 83}, 2481 (1999).

\refis{Har6} J. B. Hartle, in {\it Proceedings of
the Cornelius Lanczos International Centenary Confererence},
edited by J.D.Brown, M.T.Chu, D.C.Ellison and R.J.Plemmons
(SIAM, Philadelphia, 1994). See also eprint gr-qc/9404017.

\refis{HLM} J. B. Hartle, R. Laflamme and D. Marolf, 
\pr {\bf D51}, 7007 (1995).

\refis{JoZ} E.Joos and H.D.Zeh, {\sl Z.Phys.} {\bf B59}, 223 (1985).

\refis{Omn} R. Omn\`es, {\sl J.Stat.Phys.} {\bf 53}, 893 (1988);
{\bf 53}, 933 (1988);
{\bf 53}, 957 (1988);
{\bf 57}, 357 (1989);
{\sl Ann.Phys.} {\bf 201}, 354 (1990);
{\sl Rev.Mod.Phys.} {\bf 64}, 339 (1992).

\refis{SM} See for example, R.L.Liboff, {\it Introduction to the
Theory of Kinetic Equations} (Wiley, New York, 1969); K.Huang, {\it
Statistical Mechanics}, 2nd edition (New York, Chichester, Wiley,
1987).

\refis{Zur} See for example, 
W.Zurek, in {\it Physical Origins of Time Asymmetry},
edited by  J.J.Halliwell, J.Perez-Mercader and W.Zurek (Cambridge
University Press, Cambridge, 1994); preprint quant-ph/9805065.

\endreferences

\end